# Ashkin-Teller criticality and weak-first-order behavior of phase transition to four-fold degenerate state in two-dimensional frustrated Ising antiferromagnets

R. M. Liu[1], W. Z. Zhuo[1], J. Chen[1], M. H. Qin[1] [*], M. Zeng[1], X. B. Lu[1], X. S. Gao[1], and J. –M. Liu[2] [†]

[1]*Institute for Advanced Materials, South China Academy of Advanced Optoelectronics and Guangdong Provincial Key Laboratory of Quantum Engineering and Quantum Materials, South China Normal University, Guangzhou 510006, China*

[2]*Laboratory of Solid State Microstructures & Innovative Center of Advanced Microstructures, Nanjing University, Nanjing 210093, China*

[**Abstract**] We study the thermal phase transitions of the four-fold degenerate phases (the plaquette and single stripe states) in two-dimensional frustrated Ising model on the Shastry-Sutherland lattice using Monte Carlo simulations. The critical Ashkin-Teller-like behavior is identified both in the parameter regions with the plaquette and single stripe phases, respectively. The four-state Potts-critical end points differentiating the continuous transitions from the first-order ones are estimated based on finite-size scaling analyses. Furthermore, similar behavior of the transition to the four-fold single stripe phase is also observed in the anisotropic triangular Ising model. Thus, this work clearly demonstrates that the transitions to the four-fold degenerate states of two-dimensional Ising antiferromagnets exhibit similar transition behavior.

---

[*] qinmh@scnu.edu.cn
[†] liujm@nju.edu.cn

## I. Introduction

New ground states may arise in classical spin models when competing interactions are introduced.[1] For example, in the well-known two-dimensional (2D) square-lattice Ising model, the single stripe antiferromagnetic (AFM) state (spin configuration is shown in Fig. 1(a)) rather than the Ising AFM state (Fig. 1(b)) is stabilized at low temperatures ($T$) when the next nearest neighbor (NNN) AFM coupling $J_2$ increases to above 1/2 of the nearest neighbor (NN) AFM coupling $J_1$.[2-5] Interestingly, unlike the Ising transition to the Ising AFM state which breaks a twofold ($Z_2$) symmetry, the nature of the phase transition from a paramagnetic state to the single stripe state (a four-fold $Z_4$ symmetry is broken) is more complicated and cannot be directly determined from the symmetry of the order parameter. This subject has been extensively investigated using various methods in earlier works.[6,7]

In 1993, the mean-field calculation[8] gave evidence for a first-order transition in a certain region of the frustration $J_2/J_1$, challenging the earlier scenario[4] of a continuous stripe phase transition with varying exponents. Subsequently, the first-order scenario was supported by Monte Carlo (MC) simulations[9], and the general picture of two different transition scenarios (transitions of first-order and Ashkin-Teller (AT)-like characters) was uncovered[10] using conformal field theory and MC simulations. This picture is well accepted, and the critical end point between the two scenarios was reasonably estimated to be $J_2/J_1 \approx 0.67$ by employing a combination of MC simulations and finite-size scaling analyses.[11-13] In our earlier work, it was proven that further-neighbor interactions can also modulate the critical behaviors of the model.[14]

While the phase transitions in the square-lattice Ising model have been clarified, studies proceed in the models for other frustrated magnetic materials such as Shastry-Sutherland (S-S)[15,16] and triangular[17,18] lattice antiferromagnets. The study of the S-S Ising model becomes very important from the following two viewpoints. On one hand, this model[19,20] can be effectively used to study the magnetic properties of the rare-earth tetraborides such as TmB$_4$[16]. More importantly, for such a magnetic system of relatively low lattice symmetry, the further neighboring interactions may no longer be negligible in comparison with the NN and NNN interactions. As a matter of fact, the further neighboring interactions are believed[21-23] to be very important in modulating the fascinating magnetization behaviors experimentally reported[16] in TmB$_4$. Thus, the study of the phase transitions in the S-S Ising model may provide useful information in understanding experimental observations. On the other hand, this study also contributes to the development of statistical mechanics and solid-state physics. For example, besides the Ising AFM (Fig. 2(a)) and single stripe (Fig. 2(b)) states mentioned above, two four-fold degenerate plaquette states (Fig. 2(c) and Fig. 2(d)) can be developed in certain parameter regions for the S-S Ising model, respectively. However, it is still an open question whether these phase-transition scenarios introduced above (for the stripe phase transition) hold true for the transitions between the plaquette state and a disordered state.

Furthermore, the 2D triangular-lattice Ising model[24-27] has been successfully used to study the magnetization behaviors[28] in the spin-chain system Ca$_3$Co$_2$O$_6$. In the model with the NN and NNN AFM couplings, a six-fold degenerate single stripe state is developed at low $T$.[29,30] Earlier analysis has shown that the destruction of the state has to take place via a phase transition of first-order character.[31,32] The AFM coupling $J_2$ only modulates the transition point, and cannot alter the transition character. However, when a spatial anisotropy is introduced, the six-fold degeneracy of the single stripe state is decreased to a four-fold one, as will be systematically discussed in Section III. Naturally, one may question that if the transition character is also altered in the anisotropic model. In fact, the question is related to the universality of these transitions, and definitely deserves to be checked in detail.

In this work, we study the frustrated Ising model on the S-S lattice using MC simulations, and pay particular attention to critical behaviors of the phase transitions to the four-fold degenerate states (the plaquette and single stripe orders). The critical behaviors are revealed to be similar to those of the square-lattice Ising model: weak first-order transition, four-state Potts criticality and AT-like critical behavior (the critical exponents vary continuously between those of the four-state Potts model and the Ising model). Furthermore, similar phase transition behaviors of the anisotropic triangular Ising model are also observed, suggesting that the transitions to four-fold degenerate states may be of a similar universality.

The rest of this paper is organized as follows. In Sec. II we introduce the S-S Ising model and present the simulated results. Section III is dedicated to the study of the anisotropic triangular Ising model. The conclusion is presented in Sec. IV.

## II. Phase transitions in the frustrated Ising model on the S-S lattice

In this section, we study the phase transitions of the S-S $J_1$-$J_2$-$J_3$ Ising model which can be written by the classical Hamiltonian

$$H = J_1 \sum_{\langle ij \rangle_1} S_i S_j + J_2 \sum_{\langle ij \rangle_2} S_i S_j + J_3 \sum_{\langle ij \rangle_3} S_i S_j, \qquad (1)$$

where $J_n$ are the exchange couplings between the $n$-th nearest neighbor spins $\langle ij \rangle_n$ as shown in Fig. 2(a), and $S_i = \pm 1$ is the Ising spin on site $i$. The Ising AFM order (Fig. 2(a)), the single stripe order (Fig. 2(b)), and the two plaquette orders (Fig. 2(c) and 2(d)) can be stabilized at low $T$ in different exchange parameter regions, respectively. To simplify the discussion on the relevant phases, we use $J_2 = 1$ as the energy unit, and take other couplings as variables. Our simulation is performed using the standard Metropolis algorithm and the parallel tempering algorithm.[33,34] We take an exchange sampling after every 10 standard MC steps. Generally, the initial $5\times10^5$ MC steps are discarded for equilibrium consideration and another $5\times10^5$ MC steps are retained for statistical averaging of the simulation.

### A. Phase transition to the plaquette state from the high-T paramagnetic order

The ground-state phase diagram of the model can be reasonably obtained by comparing the energies of the four possible phases.[35] For example, the plaquette state (Fig. 2(c)) occupies the region of $J_1 > 2$ and $J_3 < 0$. For a description of the plaquette order, the order parameter $m_p$ can be similarly defined as[11]

$$m_p^2 = m_1^2 + m_2^2, \qquad (2)$$

with

$$m_1 = \frac{1}{N} \sum_i A_1^i S_i, \quad m_2 = \frac{1}{N} \sum_i A_2^i S_i, \qquad (3)$$

where the value of $A^i$ 1 depends on the coordinates of site $i$ on an $N = L \times L$ ($24 \leq L \leq 256$) periodic lattice and the spin configuration shown in Fig. 2(c) (Specifically, $A^i$ 1 = 1 for up-spin site, and $A_{i1} = -1$ for down-spin site), and $A_{i2}$ depends on the other spin configuration (can be obtained by rotating the configuration in Fig. 2(c) by 90 degree). Furthermore, in order to study the critical behaviors of the phase transition, we calculate the susceptibility $\chi_p$:

$$\chi_p = \frac{N\left(\langle m_p^2 \rangle - \langle |m_p| \rangle^2\right)}{T}, \tag{4}$$

and the Binder cumulant $U_p$:

$$U_p = 2\left(1 - \frac{1}{2}\frac{\langle m_p^4 \rangle}{\langle m_p^2 \rangle^2}\right), \tag{5}$$

where $\langle \ldots \rangle$ is the ensemble average.

Fig. 3(a) shows the calculated $U_p$ as a function of $T$ for various $L$ at $J_1 = 2.02$ and $J_3 = -0.7$. A negative peak is developed even for the smallest system size $L = 24$ and grows with increasing $L$, strongly suggesting a first-order transition between the plaquette state and a high-$T$ disordered state.[36] Similarly, we also calculate the scaling exponent $\gamma/\nu$ which is estimated from the local slope of the peak values of $\chi_p$ ($\chi_{\max}$, given in the inset of Fig. 3(b)) between $L$ and $L/2$. The exponent $\gamma/\nu$ for various $L$ is shown in Fig. 3(b), which seems to be converged to about 1.82 for $L \geq 128$. It is noted that $\gamma/\nu$ should approach 2 at infinite $L$ for a first-order transition. However, similar evolutions of $\gamma/\nu$ with $L$ have been reported in the square-lattice $J_1$-$J_2$ and AT models, which are attributed to the large correlation length related to a very weak first-order transition.[13] Thus, the plaquette AFM transition for small $J_1 = 2.02$ and $J_3 = -0.7$ is also of weak first-order character.

The first-order transition can be modulated into a continuous one as $J_1$ increases. In detail, the system size needed to develop a negative peak in $U_p$-$T$ curve increases with the increasing $J_1$, indicating an enhancing continuity of $m_p$. For example, no negative peak can be observed even for the largest $L = 256$ at $J_1 = 2.5$ as shown in Fig. 3(c), suggesting a second-order transition between the plaquette state and a disordered state.[37] This behavior can be further confirmed from the flowgram of the susceptibility at the transition temperature $T_C$ for various

$J_1$ at $J_3 = -0.7$, as presented in Fig. 3(d). The flow for a fixed $J_1$ ($J_1 \geq 2.15$) almost equals to a constant in the scaling over $L^{-7/4}$, exhibiting the AT critical behavior.[38] However, the flow at $J_1 = 2.02$ changes a lot with $L$. The significant change of the behavior of the flow of the susceptibility strongly suggests that the type of this transition is altered, as has been pointed out in earlier works.[39]

In the parameter region with the AT critical behavior, the critical exponents should be varied with the magnitude of the frustration (the ratio of $J_1/J_3$ to $J_2$). This phenomenon is also confirmed in our simulations, i.e., the estimated $v$ increases with the increasing $J_1$. For example, for $J_3 = -0.7$, $v = 0.71(5)$ at $J_1 = 2.1$ ($T_C = 1.117(5)$) and $v = 0.82(2)$ at $J_1 = 2.5$ ($T_C = 1.302(2)$) are estimated. In Fig. 4(a) and 4(b), we plot the simulated $U_p$ in the scaling form: $U_p = f(tL^{1/v})$ with $t = (T - T_C)/T_C$ at $J_1 = 2.1$ and 2.5, respectively. In a wide lattice-size regime, these $U_p$ curves are well coincident with each other, confirming the estimated values of $v$. Furthermore, the existence of the decoupled Ising limit is also demonstrated at a rather large $J_1$. For example, the transition at $J_1 = 5$ is with $v \sim 1$, as revealed in the insert of Fig. 4(b), exhibiting the 2D Ising transition behavior. It is noted that the XY model in a four-fold anisotropy field is also with exponents varying with the magnitude of the field, i.e., $\beta > 1/8$ is reported under the anisotropy field.[40] However, this scenario is clearly ruled out by the estimated exponents ($1/12 < \beta < 1/8$ is confirmed in this parameter region), although the corresponding results are not shown here for brevity.

Following the earlier works, the map between the AT and S-S critical points can also be established by analyzing the universality of the Binder cumulants.[13] Fig. 4(c) shows the Binder cumulant crossing points for pairs ($L$, $L/2$) and infinite $L$ extrapolated $U_p^*$ (the curves are fitted by $U = a + b/L^c$) for $J_1 = 2.1$ and $J_1 = 2.5$ at $J_3 = -0.7$. In this case, $U_p^*$ increases monotonically with the increasing $J_1$. It is revealed that the phase transition at $J_1 = 2.1$ ($U_p^* \approx 0.815$) should map to $K \approx 0.85$ ($K$ is the coupling between the two Ising variables of the square-lattice AT model) and $J_1 = 2.5$ ($U_p^* \approx 0.873$) to $K \approx 0.4$. Here, the Binder cumulant crossing points for the AT model are not repetitively given, and one can find these values in previous publication[13]. The peak values of the specific heat $C_{max}$ for various $L$ for the two models at these points are given in Fig. 4(d), and the critical exponent $\alpha/v$ can be estimated by standard finite-size scaling argument, $C_{max} \sim L^{\alpha/v}$. It is clearly shown that $\alpha/v$ for the two models converges to the same value in a similar way at the corresponding $J_1$ and $K$, strongly convincing the map between the parameters of the AT model and the S-S Ising model.

As a matter of fact, the energy ($E$) histograms have been calculated in our simulations, and one example is shown in Fig. 5(a) which gives the results at $J_1 = 2.07$ and $J_3 = 0.7$. Two delta-functions in the energy histogram are observed for small lattice sizes ($L \leq 128$, at least), while single peak of the histogram is exhibited for large enough sizes ($L \geq 256$), clearly demonstrating a pseudo-first-order behavior of the transition. The genuine transition behavior only can be observed for systems with large sizes, similar to that of the frustrated square-lattice model.

Thus, we do finite-size scaling analyses to reasonably estimate the boundary between the two scenarios (first-order and AT-like characters), following earlier works. The Binder cumulant crossing points and $U_p^*$ for $J_1 = 2.02$ and $J_1 = 2.07$ at $J_3 = -0.7$ are shown in Fig. 5(b). Finally, the transition point between the two phase-transition scenarios in the S-S Ising model at $J_3 = -0.7$ is approximately estimated to be $[J_1]_C = 2.05 \pm 0.01$ by comparing $U_p^*$ and $U^*$ for the four-state Potts model ($\approx 0.79$), as clearly shown in Fig. 5(c).

### B. Phase transition to the single stripe state from high-T paramagnetic order

Subsequently, we study the transition to the single stripe state in the S-S $J_1$-$J_2$-$J_3$ Ising model to check if these two phase-transition scenarios are also available. Specifically, the single stripe state is stabilized at low $T$ for $J_1 + 2J_3 > 2$ with $J_1 > 0$ and $J_3 > 0$. The order parameter $m_{S1}$ is defined as[11]

$$m_{S1}^2 = m_x^2 + m_y^2, \tag{6}$$

with

$$m_{x,y} = \frac{1}{N}\sum_i (-1)^{i_{x,y}} S_i. \tag{7}$$

where ($i_x$, $i_y$) are the coordinates of site $i$. Furthermore, the Binder cumulant $U_{S1}$ and susceptibility $\chi_{S1}$ are also calculated.

Actually, the two phase-transition scenarios are also verified, similar to earlier reports on the square-lattice system. As clearly shown in Fig. 6(a), $U_{S1}(T)$ indeed develops a negative peak for $L = 24$ at $J_1 = 1.2$ and $J_3 = 0.7$. With the increase of $L$, the negative peak grows and becomes narrower, strongly suggesting a first-order phase transition. In addition, the single stripe state can be stabilized by the AFM $J_1$, resulting in the increasing $T_C$ when $J_1$ is further increased. Furthermore, the system size needed to stabilize a negative peak is increased with the increase of $J_1$ and/or $T_C$, demonstrating a weakening discontinuity of $m_{S1}$. The simulated $U_{S1}(T)$ for various $L$ at $J_1 = 2.5$ and $J_3 = 0.7$ are given in Fig. 6(b), which exhibits no negative peak even for $L = 256$, indicating a continuous transition.

As a matter of fact, both the AT criticality and first-order transition are uncovered in the parameter region with the single stripe ground-state using exactly the same methods as above. Here, we do not give the corresponding results repetitively. Thus, it is strongly suggested that the two phase-transition scenarios observed in the frustrated square-lattice Ising model are also available in other systems with the four-fold degenerate single stripe ground-state.

### C. Phase diagram of the S-S $J_1$-$J_2$-$J_3$ Ising model

Other values of $J_3$ are also simulated. As a summary, the simulated phase diagram is presented in Fig. 7, in which the ground-state boundaries among different phases are obtained by comparing the energies. Particular attentions are paid to the critical behaviors of different phase transitions. The estimated critical exponent $v$ is also presented, whereas the transition temperature $T_C$ is not given here.

On one hand, the phase transition to the Ising AFM order (The upper left corner in the phase diagram) belongs to the 2D Ising universality class. As a matter of fact, the S-S $J_1$-$J_2$ Ising model has been solved exactly, and was proved to fall into the same universality class.[41] On the other hand, the phase transition from the paramagnetic state to the plaquette state or the single stripe state exhibits different transition behaviors (first-order and AT-like characters) depending on the parameters. Specifically, in the region $J_1 \geq [J_1]_C$ (the black circles in Fig. 7) at a fixed $J_3$, the critical exponent $v$ varies continuously between that of the four-state Potts model ($v = 2/3$) at $[J_1]_C$ and the Ising model ($v = 1$) for infinite $J_1$, demonstrating the AT criticality, as clearly shown in Fig. 7. Furthermore, in the small region $J_1 < [J_1]_C$ at a fixed $J_3$ with the plaquette/single stripe state, the phase transition is of weak first-order character, as revealed in our simulations.

So far, our work clearly demonstrates that the nature of the phase transition between a four-fold degenerate state and a disordered state in the 2D S-S Ising model depends on the values of the frustration ($J_1/J_2$ and/or $J_3/J_2$), similar to that of the stripe phase transition in the square-lattice Ising model. The simulated results may be experimentally realized in the rare-earth tetraborides such as $TmB_4$ and $ErB_4$ which are strong Ising magnets.[42] As a matter of fact, the third-neighbor interaction $J_3$ is expected to be negative in the $ErB_4$ compound, and the plaquette phase is supposed to be the ground state at zero magnetic field, related to those predicted in earlier theoretical calculations.[22,43] Furthermore, additional AFM $J_3$ is available in $TmB_4$ which may be with the single stripe ground-state.[20] The exchange interactions could be efficiently modulated by various experimental methods such as applying uniaxial strain and/or ion doping in these systems. Thus, the critical temperatures and the characters of the phase transitions may be altered in the strained and/or the doped systems, as predicted in our simulations.

## III. Phase transitions in the anisotropic triangular Ising antiferromagnet

In this section, we study the phase transitions of an anisotropic AFM Ising model on the triangular lattice, in order to check the universality of the phase transition to a four-fold degenerate state. The model Hamiltonian can be written as

$$H = \sum_{\langle ij \rangle_1} J_{ij} S_i S_j + J_2 \sum_{\langle ij \rangle_2} S_i S_j. \tag{8}$$

here, a spatial anisotropy modulated by $dJ_1$ (the NN coupling $J_1 + dJ_1$ along one of the three directions, solid red lines in Fig. 8(a)) is considered. We fix $J_1 = 1$ and change $dJ_1$ and the NNN coupling $J_2$ to study the phase transition behaviors. The order parameter of the four-fold degenerate single stripe state (two ground-state structures are shown in Fig. 8, the other structures can be obtained through reversing all the spins) $m_{S2}$ and the Binder cumulant $U_{S2}$ are calculated by similar equations as above.

Interestingly, the transition character can also be modulated by $J_2$ even when a small anisotropy $dJ_1 = 0.05$ is introduced in the model. Fig. 9(a) shows the calculated $U_{S2}$ as a function of $T$ for various $L$ at $J_2 = 0.1$, which exhibits a negative peak for $L = 24$, demonstrating an obvious first-order transition. All the negative peaks disappeared when $J_2$ increases to above 0.3 (Fig. 9(b)), indicating an enhancing continuity. Similarly, we also estimate the four-state Potts critical end points $[J_2]_C$ and the critical exponent $v$, and the corresponding results are summarized in Fig. 10 in which the ground-state phase diagram in the $(dJ_1, J_2)$ parameter plane is presented. Two scenarios of the four-fold stripe phase transition apply depending on the value of frustration, as uncovered in our simulations. Furthermore, for a fixed $dJ_1$, the phase transition is with the AT criticality for $J_2 \geq [J_2]_C$, whereas exhibits first-order behavior for $J_2 < [J_2]_C$, as shown in Fig. 10. Thus, the work in this section also suggests that the phase transition between the four-fold degenerate state and the paramagnetic state in 2D Ising models are with a similar universality.

As a matter of fact, the isotropic triangular-lattice $J_1$-$J_2$ Ising model has been studied,[32] where it was pointed out that the cubic term in the equal height model drives the phase transition to be the first order. When a positive $dJ_1$ is taken into account, the highly triple rotational symmetry would be broken, and the cubic term in the height model is depressed. The first order character of the transition for nonzero $dJ_1$ may be qualitatively understood from the following points. The doubly degenerate single stripe state can be developed at low $T$ for a negative $dJ_1$, and its destruction takes place via a Ising phase transition. However, the difference of the free-energy minima of the doubly degenerate and four-fold degenerate single stripe phases is proportional to $dJ_1$, as will be proved in the appendix. Thus, for small positive $dJ_1$, the energy gap between the states is very small and can be overcome by thermal fluctuations near the critical point, resulting in the phase separation in the system. With the increase of $dJ_1$, the magnitude of free energy gap is increased, and the region with the first order transition is depressed, as shown in the simulated phase diagram. In addition, the first order transition always exists near the phase boundary between two successive phases in the parameter space, as revealed in the simulated phase diagrams of the square-lattice and S-S Ising models, further supporting this viewpoint.

As a matter of fact, there are some real materials such as $Ca_3Co_2O_6$ compound which can be described by the 2D triangular-lattice Ising model. In some extent, the exchange anisotropy may be realized in the triangular-lattice antiferromagnets by controlling the crystal lattice using uniaxial strain/stress applied along one of the in-plane three bonds. Thus, the four-fold degenerate stripe spin structure is supposed to emerge in the strained $Ca_3Co_2O_6$, which deserves to be checked in further experiments.

However, the present work seems to reveal once more that frustrated spin systems such as the S-S and triangular Ising antiferromagnets have different four-fold degenerate AFM states in which the thermal phase-transitions may be with similar universality. The phase transitions in these systems have attracted attention for many years, but their relations remained ambiguous before the present simulations.

## IV. Conclusion

In conclusion, we study the nature of the thermal phase transitions to four-fold degenerate phases (the plaquette state and single stripe state in the S-S Ising model, and the four-fold degenerate stripe state in the anisotropic triangular Ising model) using Monte Carlo simulations and finite-size scaling analyses. The critical Ashkin-Teller-like behaviors are revealed in the parameter regions with these orders, respectively. Furthermore, the first-order behaviors are observed below the four-state Potts-critical end points, similar to earlier reports. Thus, this work strongly suggests that the transitions to four-fold degenerate states of 2D Ising antiferromagnets may be with a similar transition behavior.

**Acknowledgements**:

The work is supported by the National Key Projects for Basic Research of China (Grant No. 2015CB921202), and the National Key Research Programs of China (Grant No. 2016YFA0300101), and the Natural Science Foundation of China (Grants No. 51332007), and the Innovation Project of Graduate School of South China Normal University, and Joint International Research Laboratory of Optical Information, and the Science and Technology Planning Project of Guangdong Province (Grant No. 2015B090927006). X. Lu also thanks for the support from the project for Guangdong Province Universities and Colleges Pearl River

Scholar Funded Scheme (2016).

# Appendix

Using the mean field approximate approach, we calculate the free energy per site of the four-fold degenerate single stripe phase by:[44]

$$f_4 = (J_1 + dJ_1 + J_2)m^2 - \frac{1}{\beta}\ln\{2\cosh[2\beta(J_1 + dJ_1 + J_2)m]\}, \tag{A1}$$

where $m$ is the average magnetization and $\beta = 1/T$. Then, we calculate the critical transition temperature $T_{4c}$, and expand the free energy with respect to $m$ at temperature $T_4 = \alpha T_{4c}$ ($\alpha < 1$ and $\sim 1$) by:

$$T_{4c} = 2(J_1 + dJ_1 + J_2), \tag{A2}$$

and

$$f_{4c} \approx -\frac{1}{\beta}\ln 2 + [(J_1 + dJ_1 + J_2) - 2\beta(J_1 + dJ_1 + J_2)^2]m^2 + \frac{4}{3}\beta^3(J_1 + dJ_1 + J_2)^4 m^4, \tag{A3}$$

repectively. According to Landau theory, the equation (A3) demonstrates a second order transition.[45] One minimizes $f_{4c}$ with respect to $m$ and obtains:

$$f_{4c}^{\min} \approx -2\alpha(J_1 + dJ_1 + J_2)\ln 2 - \frac{3}{2}\alpha(1-\alpha)^2(J_1 + dJ_1 + J_2), \tag{A4}$$

Similarly, for the doubly degenerate single stripe phase, the minimum of the free energy $f_{2c}$ at $T_2 = \alpha T_{2c}$ ($\alpha < 1$ and $\sim 1$) can be calculated by:

$$f_{2c}^{\min} \approx -2\alpha(J_1 - dJ_1 + J_2)\ln 2 - \frac{3}{2}\alpha(1-\alpha)^2(J_1 - dJ_1 + J_2), \tag{A5}$$

with

$$T_{2c} = 2(J_1 - dJ_1 + J_2), \tag{A6}$$

Thus, near the transition point, the difference of the free energy minima of the two successive phases in the parameter space is:

$$\Delta f_{\min} \approx [3\alpha(1-\alpha)^2 + 4\alpha\ln 2]dJ_1 \propto dJ_1. \tag{A7}$$

Scientific, 2005).

**FIGURE CAPTIONS**

Figure 1. (color online) Spin configurations in (a) the single stripe state, and (b) the Ising AFM state of the frustrated square-lattice $J_1$-$J_2$ Ising model. Solid and empty circles represent the up-spins and down-spins, respectively.

Figure 2. (color online) Spin configurations in (a) the Ising AFM state, and (b) the single stripe state, and (c) (d) the two plaquette states of the S-S $J_1$-$J_2$-$J_3$ Ising model.

Figure 3. (color online) Binder cumulant $U_p$ as a function of $T$ for different $L$ at $J_3 = -0.7$ at (a) $J_1 = 2.02$ and (c) $J_1 = 2.5$. (b) Scaling exponent $\gamma/\nu$ for various $L$, and the insert shows $\chi_{max}$ for various $L$, and (d) Flowgrams for the susceptibility $\chi_p$ at the transition temperature $T_C$ for various $J_1$ at $J_3 = -0.7$.

Figure 4. (color online) A scaling plot of $U_p$ at $J_3 = -0.7$ at (a) $J_1 = 2.1$ and (b) $J_1 = 2.5$ in the scaling form: $U_p = f(tL^{1/\nu})$ with $t = (T - T_C)/T_C$. The insert shows the results at $J_1 = 5$. (c) Binder cumulant crossing points $U_p$ for ($L$, $L/2$) system pairs and the extrapolation to infinite $L$ for $J_1 = 2.1$ and $J_1 = 2.5$ at $J_3 = -0.7$, and (d) the peak values of the specific heat vs $L$ for the S-S Ising and square-lattice AT models.

Figure 5. (color online) (a) Energy histograms for different $L$ at critical temperatures at $J_1=2.07$ and $J_3=-0.7$, (b) and (c) $U_p^*$ of the model for various $J_1$ at $J_3 = -0.7$ compared with that of the 4-state Potts model shown with the dotted line.

Figure 6. (color online) Binder cumulant $U_{S1}$ as a function of $T$ for different $L$ at $J_3 = 0.7$ at (a) $J_1 = 1.2$ and (b) $J_1 = 2.5$.

Figure 7. (color online) Ground-state phase diagram for the frustrated S-S $J_1$-$J_2$-$J_3$ Ising model. The estimated $\nu$ is also depicted. The critical exponents in the parameter region covered with black solid lines could not be well estimated.

Figure 8. (color online) Spin configurations in the single stripe state of the anisotropic triangular-lattice Ising model. The other two spin configurations in the four-fold degenerate plaquette state can be obtained by reversing the spins in (a) and (b), respectively. In (a), the exchange couplings are also shown.

Figure 9. (color online) Binder cumulant $U_{S2}$ as a function of $T$ for various $L$ at $dJ_1 = 0.05$ at (a) $J_2 = 0.1$ and (b) $J_2 = 0.3$.

Figure 10. (color online) Ground-state phase diagram in the $(dJ_1, J_2)$ space for the anisotropic Ising model on the triangular lattice. The estimated $v$ is also depicted.

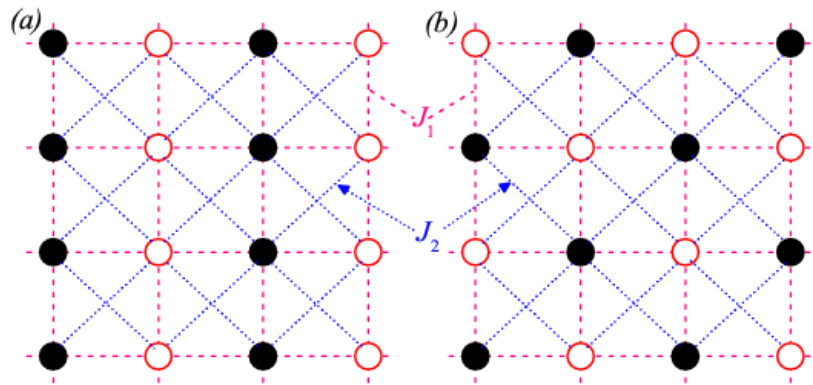

Figure 1. (color online) Spin configurations in (a) the single stripe state, and (b) the Ising AFM state of the frustrated square-lattice $J_1$-$J_2$ Ising model. Solid and empty circles represent the up-spins and down-spins, respectively.

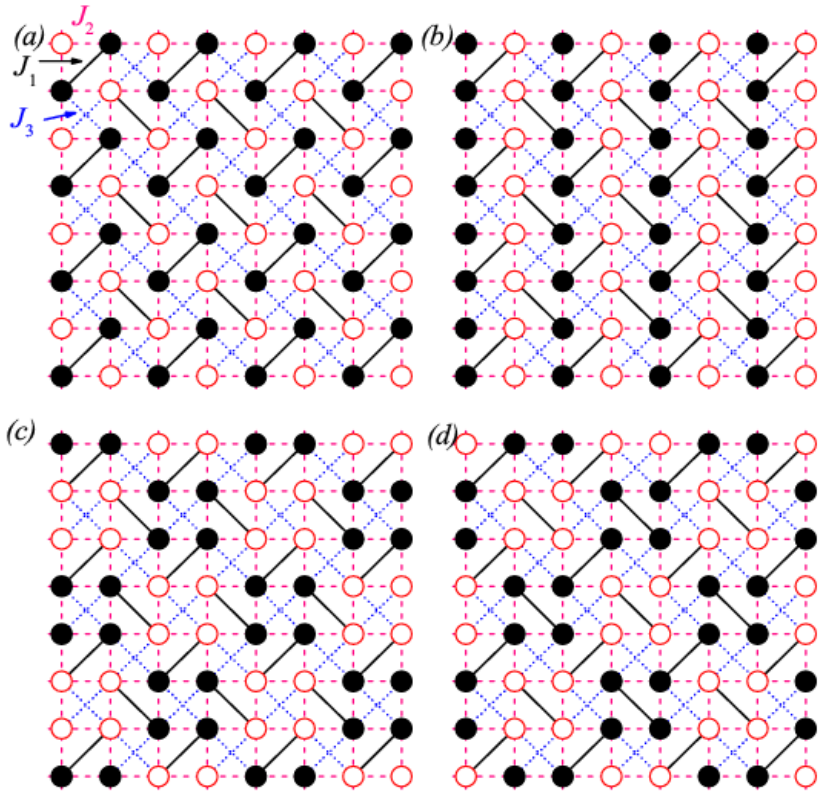

Figure 2. (color online) Spin configurations in (a) the Ising AFM state, and (b) the single stripe state, and (c) (d) the two plaquette states of the S-S $J_1$-$J_2$-$J_3$ Ising model.

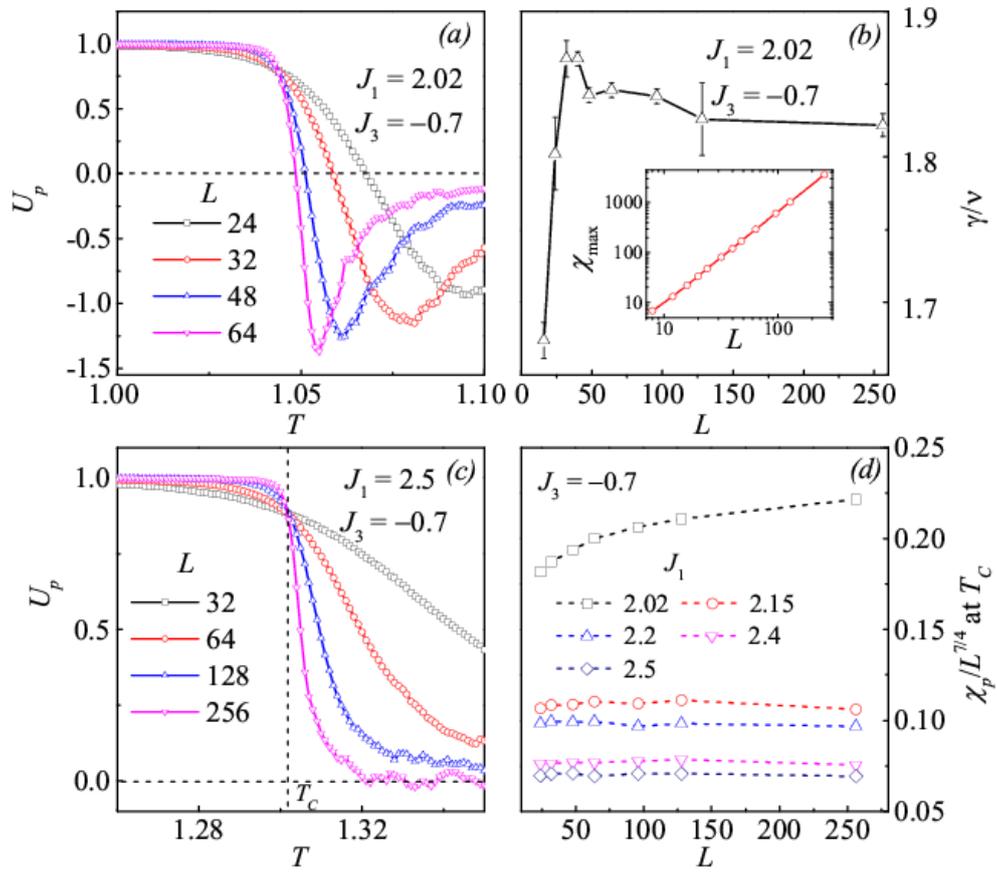

Figure 3. (color online) Binder cumulant $U_p$ as a function of $T$ for different $L$ at $J_3 = -0.7$ at (a) $J_1 = 2.02$ and (c) $J_1 = 2.5$. (b) Scaling exponent $\gamma/\nu$ for various $L$, and the insert shows $\chi_{max}$ for various $L$, and (d) Flowgrams for the susceptibility $\chi_p$ at the transition temperature $T_C$ for various $J_1$ at $J_3 = -0.7$.

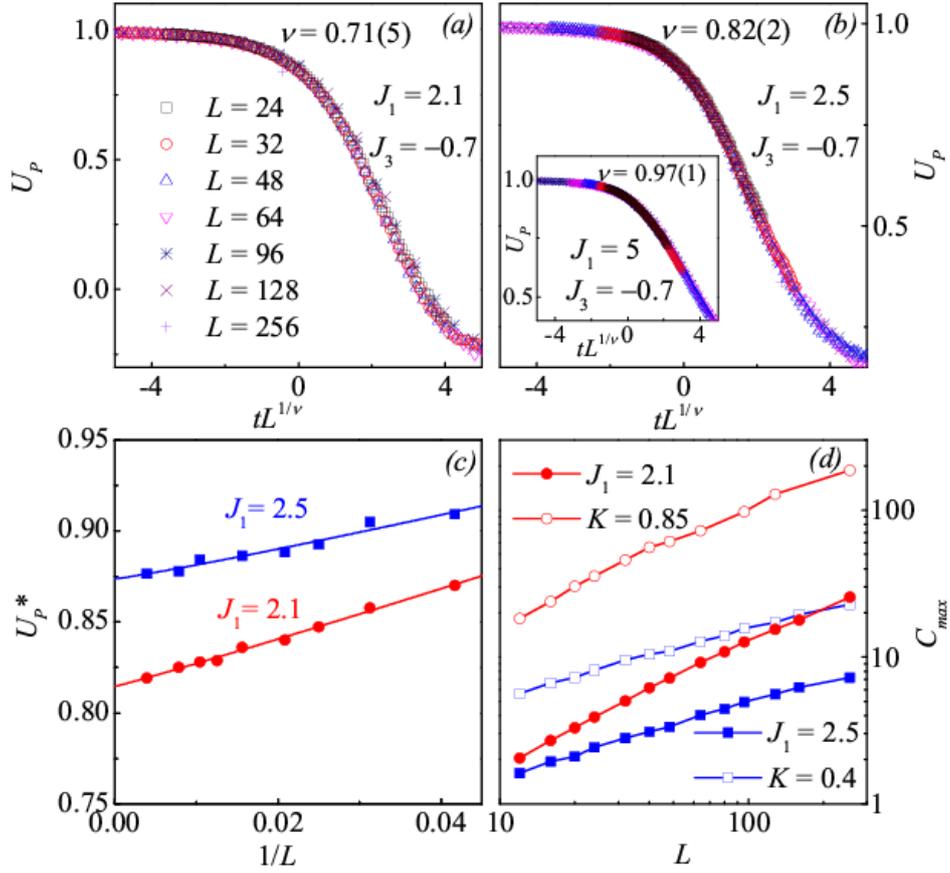

Figure 4. (color online) A scaling plot of $U_p$ at $J_3 = -0.7$ at (a) $J_1 = 2.1$ and (b) $J_1 = 2.5$ in the scaling form: $U_p = f(tL^{1/\nu})$ with $t = (T - T_C)/T_C$. The insert shows the results at $J_1 = 5$. (c) Binder cumulant crossing points $U_p$ for $(L, L/2)$ system pairs and the extrapolation to infinite $L$ for $J_1 = 2.1$ and $J_1 = 2.5$ at $J_3 = -0.7$, and (d) the peak values of the specific heat vs $L$ for the S-S Ising and square-lattice AT models.

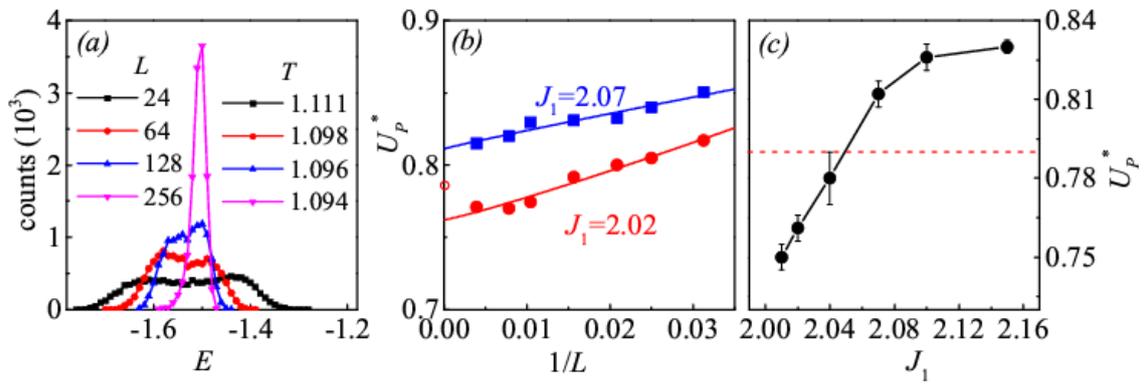

Figure 5. (color online) (a) Energy histograms for different $L$ at critical temperatures at $J_1=2.07$ and $J_3=-0.7$, (b) and (c) $U_P^*$ of the model for various $J_1$ at $J_3 = -0.7$ compared with that of the 4-state Potts model shown with the dotted line.

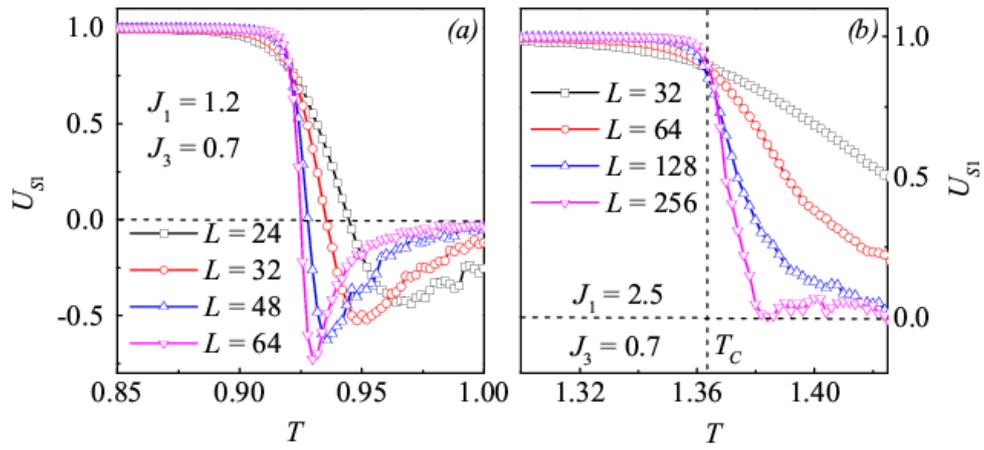

Figure 6. (color online) Binder cumulant $U_{S1}$ as a function of $T$ for different $L$ at $J_3 = 0.7$ at (a) $J_1 = 1.2$ and (b) $J_1 = 2.5$.

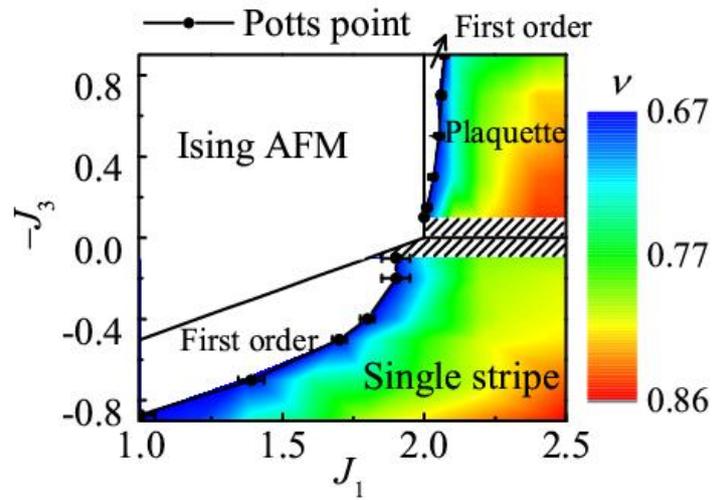

Figure 7. (color online) Ground-state phase diagram for the frustrated S-S $J_1$-$J_2$-$J_3$ Ising model. The estimated $v$ is also depicted. The critical exponents in the parameter region covered with black solid lines could not be well estimated.

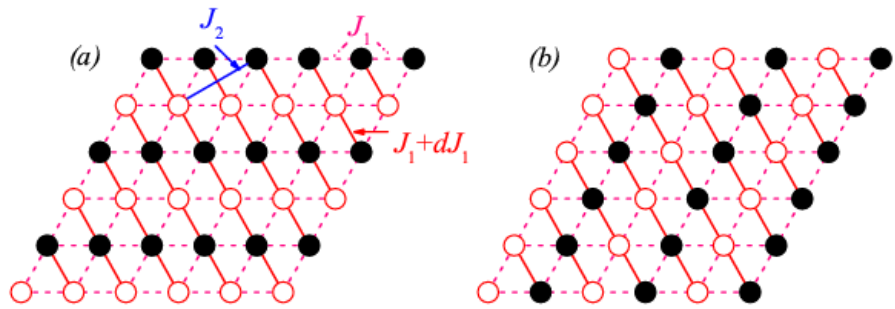

Figure 8. (color online) Spin configurations in the single stripe state of the anisotropic triangular-lattice Ising model. The other two spin configurations in the four-fold degenerate plaquette state can be obtained by reversing the spins in (a) and (b), respectively. In (a), the exchange couplings are also shown.

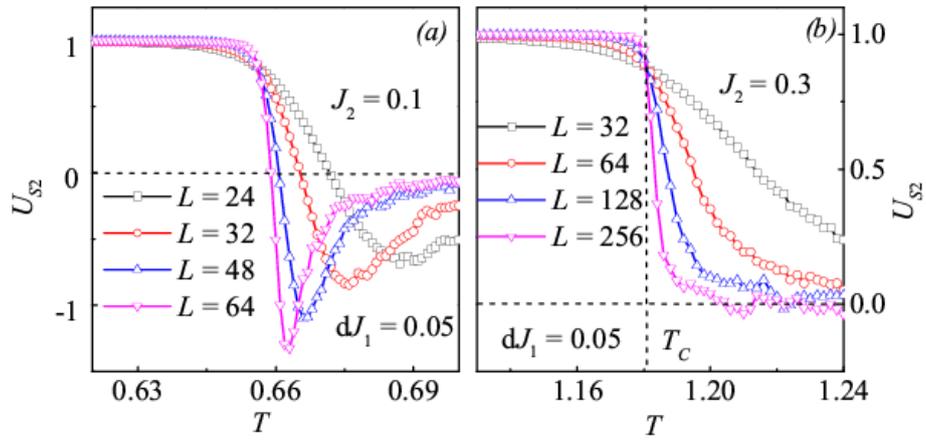

Figure 9. (color online) Binder cumulant $U_{S2}$ as a function of $T$ for various $L$ at $dJ_1 = 0.05$ at (a) $J_2 = 0.1$ and (b) $J_2 = 0.3$.

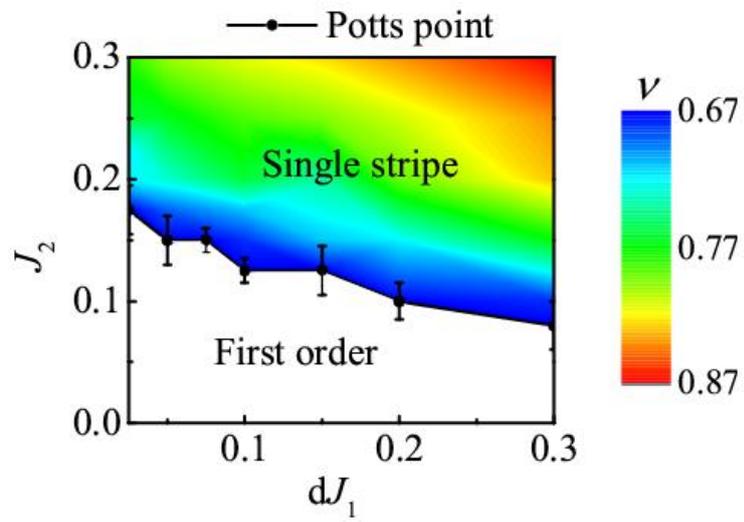

Figure 10. (color online) Ground-state phase diagram in the (d$J_1$, $J_2$) space for the anisotropic Ising model on the triangular lattice. The estimated $v$ is also depicted.